# Discontinuities in Citation Relations among Journals: Self-organized Criticality as a Model of Scientific Revolutions and Change

Loet Leydesdorff,*[a] Caroline S. Wagner,[b] and Lutz Bornmann[c]

**Abstract**

Using three-year moving averages of the complete *Journal Citation Reports* 1994-2016 of the *Science Citation Index* and the *Social Sciences Citation Index* (combined), we analyze links between citing and cited journals in terms of (1) whether discontinuities among the networks of consecutive years have occurred; (2) are these discontinuities relatively isolated or networked? (3) Can these discontinuities be used as indicators of novelty, change, and innovation in the sciences? We examine each of the $N^2$ links among the $N$ journals across the years. We find power-laws for the top 10,000 instances of change, which we suggest interpreting in terms of "self-organized criticality": co-evolutions of avalanches in aggregated citation relations and meta-stable states in the knowledge base can be expected to drive the sciences towards the edges of chaos. The flux of journal-journal citations in new manuscripts may generate an avalanche in the meta-stable networks, but one can expect the effects to remain local (for example, within a specialty). The avalanches can be of any size; they reorient the relevant citation environments by inducing a rewrite of history in the affected partitions.

**Keywords**: journal, citation, discontinuity, avalanches, criticality.

[a] *corresponding author; Amsterdam School of Communication Research (ASCoR), University of Amsterdam PO Box 15793, 1001 NG Amsterdam, The Netherlands; loet@leydesdorff.net

[b] John Glenn College of Public Affairs, The Ohio State University, Columbus, Ohio, USA, 43210; wagner.911@osu.edu

[c] Division for Science and Innovation Studies, Administrative Headquarters of the Max Planck Society, Hofgartenstr. 8, 80539 Munich, Germany; bornmann@gv.mpg.de

## 1. Introduction

With the complete *Journal Citation Reports* during the period 1994-2016 as data, we address the question of change and stability in the sciences at the level of the ($n^2$) aggregated citation links between ($n$) journals using entropy statistics. Entropy statistics allows for dynamic analysis and aggregation. This operationalization enables us to relate micro-developments in the data to theorizing about the sciences in terms of distributed change (Price, 1976; cf. Kuhn, 1962). Our results suggest that the dynamics can be explained by considering Bak *et al*.'s (1987) model of "self-organized criticality": the knowledge base can be considered as a large set of meta-stable constructs which are continuously disturbed by new knowledge claims bringing also new citation relations. "Avalanches" of variable size can then be expected. The effects, however, are local; the meta-stable regions operate in parallel. The overall system remains tending towards meta-stability at "the edge of chaos" because of the ongoing flux of new manuscripts creating and rewriting journal-journal relations in terms of citations at different scales (Zitt *et al*., 2005).

Aggregated journal-journal citation relations provide a possible baseline for studying the structural dynamics of the sciences (Studer & Chubin, 1980, p. 269). Citations and co-citations at the paper level have a high turnover in fields with a research front such as biomedicine, but less so in other fields (Garfield, 1979a; Price, 1970). Repertoires of words and co-words can be "translated" among fields with flexible interpretations (Callon, Law, & Rip, 1986). However, scholarly journals are established institutions providing quality control by peer review. This "journal literature" can be considered as the core of scientific literature (Price, 1965); new disciplinary developments can be expected to lead to new journals (Price, 1961).

Journals play a crucial and institutionalized role in the validation of knowledge claims and in the incorporation of new knowledge into the archive of science. Given their role in the codification of knowledge, journals can be considered as an organizing layer of the scientific literature. Not incidentally, the *Science Citation Index* (SCI) and its derivates (the *Social Sciences Citation Index* (SSCI) and the *Arts & Humanities Citation Index* (AHCI)) were defined in terms of specific journal selections (Garfield, 1972; 1979b), as is *Scopus*, the main competitor of the *SCI* since 2004.

The gate-keeping role of journals is nowadays debated (Brembs, Button, & Munafò, 2013; Kling & Callahan, 2003; cf. Zsindely, Schubert, & Braun, 1982). The internet revolution may have changed the landscape. Google Scholar (since 2004) collects articles on a case-by-case basis by crawling the web. Consequently, Google Scholar does not delineate the universe of scholarly documents. Furthermore, with the introduction of *PLOS ONE* in 2006, new journals have emerged that deliberately abstain from disciplinary criteria in the peer-review process in favor of a focus on novelty in terms of methods and data. As a consequence, journals may have lost some of their exclusiveness and perhaps precision in maintaining borders among disciplines (Harzing & Alakangas, 2016). Boyack & Klavans (2011; Klavans & Boyack, 2017), for example, argue that "direct citation" at the article level (Waltman & van Eck, 2012) has become an organizer of the literature more strongly than journals or other possible groupings of citations.

In this study, we use aggregated journal-journal citation *relations* as units of analysis instead of journals. Each relation specifically combines *two* journals. The journal-journal citation relation is a link in the networks of which journals are the nodes. The citation relations among the 10,000+ journals contained in the *Journal Citation Reports* (*JCR*) of the *Science Citation Index* and the *Social Sciences Citation Index* can be organized as a matrix of $(10{,}000+)^2$ cells, each representing a unique relation between a citing and a cited journal. These (valued) relations can change over time to the extent that they can disappear or emerge; that is, turn from a zero into a positive value larger than or equal to one.[4]

From case-study research, we know that specific journal-journal relations can catalyze structural changes in the journal networks over time. For example, Leydesdorff (1986) has shown how the introduction of the *Scandinavian Journal of Work Environment and Health*—owned by the Swedish trade-unions—triggered a clustering in the (relatively small) network of journals about occupational health around 1978 (Granovetter, 1973). Van den Besselaar & Leydesdorff (1996) found the merging of three journals (*Artificial Intelligence*, *AI Magazine*, and *IEEE Expert*)[5] into a citation community decisive for the take-off of artificial intelligence as a specialism since 1988. Major disciplinary developments such as the development of nanoscience and technology in recent decades can also be analyzed and visualized in terms of journal citation relations (e.g., Leydesdorff & Schank, 2008; Rosvall & Bergstrom, 2010; cf. Klavans & Boyack, 2009). However, new developments at the level of specialties and disciplines cannot be indicated unequivocally in terms of new journals, since journals often emerge within existing fields, and then do not indicate change but growth and stability within disciplinary boundaries (Leydesdorff, Cozzens, & Van den Besselaar, 1994).

[4] Citation relations between journals are counted in the *JCR*s as the sum of unique citation relations among papers (Garfield, 1979).
[5] The journal name was changed to *IEEE Intelligent Systems* in 1997.

Journal names can also be changed indicating a change in focus or, for example, reflecting a shift from a national to an international orientation. Each year, the two *JCRs* (of the SCI and SSCI) provide lists of the changes in the previous year.[6] Note that changes in journal names and/or journal citation relations do not indicate novelty at the level of individual papers (Uzzi, Mukherjee, Stringer, & Jones, 2013; Wang, Veugelers, & Stephan, 2017). From an evolutionary perspective, atypical citation patterns in papers provide the variation, whereas journal structures can be considered as selection environments. Changes at the aggregated level are more structural and occur more slowly than those at the level of articles containing knowledge claims. We will analyze the asymmetrical journal-journal citation matrix from the "citing" side which follows the research front, while "cited" represents the archive that can be expected to develop at a lower speed (Leydesdorff, 1995; Small, 1978; Small & Griffith, 1974)).

In sum, we compare cells within citation matrices of the aggregated journal-journal citation relations during the period 1994-2016 taking journal name changes into account. 1994 is the first year that the *JCR*s were available as CD-Roms (replacing earlier paper and microfiche versions) and 2016 is the last year for which this data is available. The comparison is done using (1) dynamic measures such as Kullback & Leibler's (1951) divergence measure (KL), which enables us to compare the relative frequency distribution *ex ante* as an expectation of the distribution *ex post*; (2) Theil's (1972) measure of improving on or worsening the prediction using KL divergences; and (3) Leydesdorff's (1991) measure of critical path-dependency based on Theil's measure. (The analytical relations are elaborated in the Appendices.)

---

[6] For analytical reasons, one may wish to include only journals present during all the years under study or, in other words, a "fixed journal set" (Narin, 1976; cf. Leydesdorff, 1988).

The use of algorithms that focus on the dynamics represents an improvement over comparing statics maps year-by-year consecutively (e.g., Rosvall & Bergstrom, 2010). The dynamics is more than the resulting difference between two states. Because we focus on journal-journal links, those that pass the statistical tests can be recomposed into networks and analyzed and visualized by standard programs such as Pajek and VOSviewer. Links may be connected into large components. However, incidental changes in links between *two* nodes in a single year can be the effect of chance processes and should therefore not be considered as equally valid indicators of change (Bianconi, Darst, Iacovacci, & Fortunato, 2014; de Nooy & Leydesdorff, 2015).

## 2. Data

The *Science Citation Index* has existed since 1964,[7] but the first edition of the *Journal Citation Reports* dates from 1975. The *Social Sciences Citation Index* was first published in 1973, and extended with *Journal Citation Reports* in 1978 (Garfield, 1979, at p. 16; Wouters, 1999). At the time, *JCRs* were made available in print and since approximately 1990 also on microfiches. Electronic versions on CD-Rom were distributed between 1994 and 2008. In 2009, this modality was abandoned in favor of downloadable files on the internet. (The internet version at WoS goes back to 1997.) However, the complete electronically available series between 1994 and 2016 is organized similarly (using MS Access), albeit with some reorganization and further extension in 2001.

[7] An experimental version of the SCI—the *Genetics Citation Index*—was available for 1961 (Wouters, 1999).

Journal coverage has been expanded over the years, with a deliberate discontinuity in 2008-2009 when the database was "regionally expanded" given its traditional underrepresentation of Eastern-European journals (Testa, 2010) and perhaps also in response to competition from *Scopus*, which was introduced in 2004 by Elsevier. Scopus covers many more journals than WoS (Leydesdorff, de Moya-Anegón, & de Nooy, 2016).

### *2.1. Descriptives*

On issuance of the *JCRs* 2016 in November 2017, one of us reorganized the *JCR* data for SCI and SSCI combined during the period 1994-2016 into a single standard format in order to obtain a cube of data containing 23 yearly slices. For each year, one can construct a matrix of *N* journals in the columns citing the same *N* journals in the rows. The number of journals *N* grows from 5,765 journals in 1994 to 11,487 in 2016 (Figure 1). Note that not all journals covered are also processed from the citing side; for example, 189 of the journals in 2016 (1.6%) were not processed for citing; articles in some journals (e.g., *The Scientist*) do not have reference lists. In our design, such cases lead to an empty column (zeros) in the matrix, but in the final analysis, we will use only values above a threshold. The zeros can thus be considered as missing values.

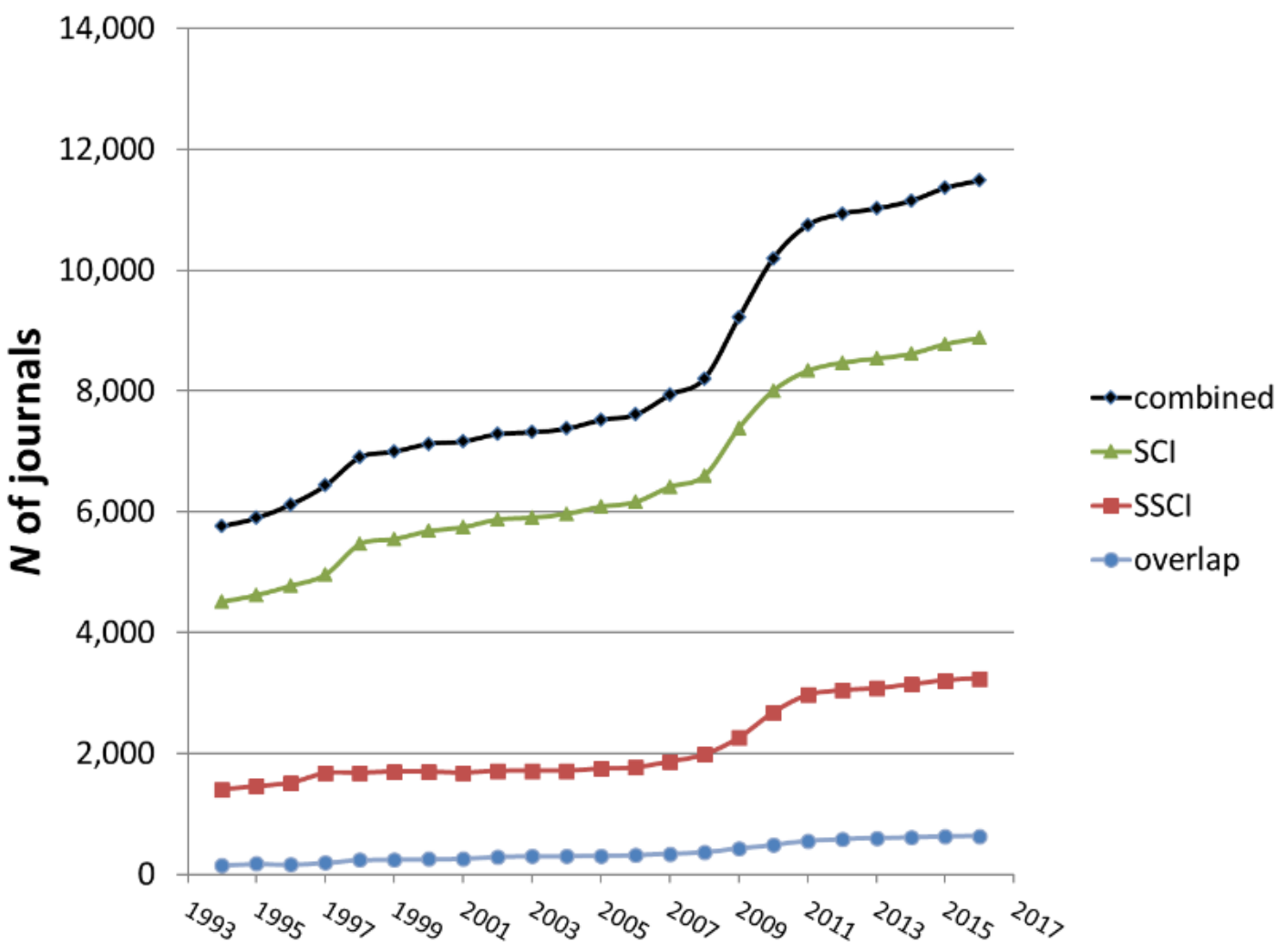

**Figure 1**: Numbers of journals in WoS for the *Science Citation Index* and the *Social Sciences Citation Index*, both separate and combined.

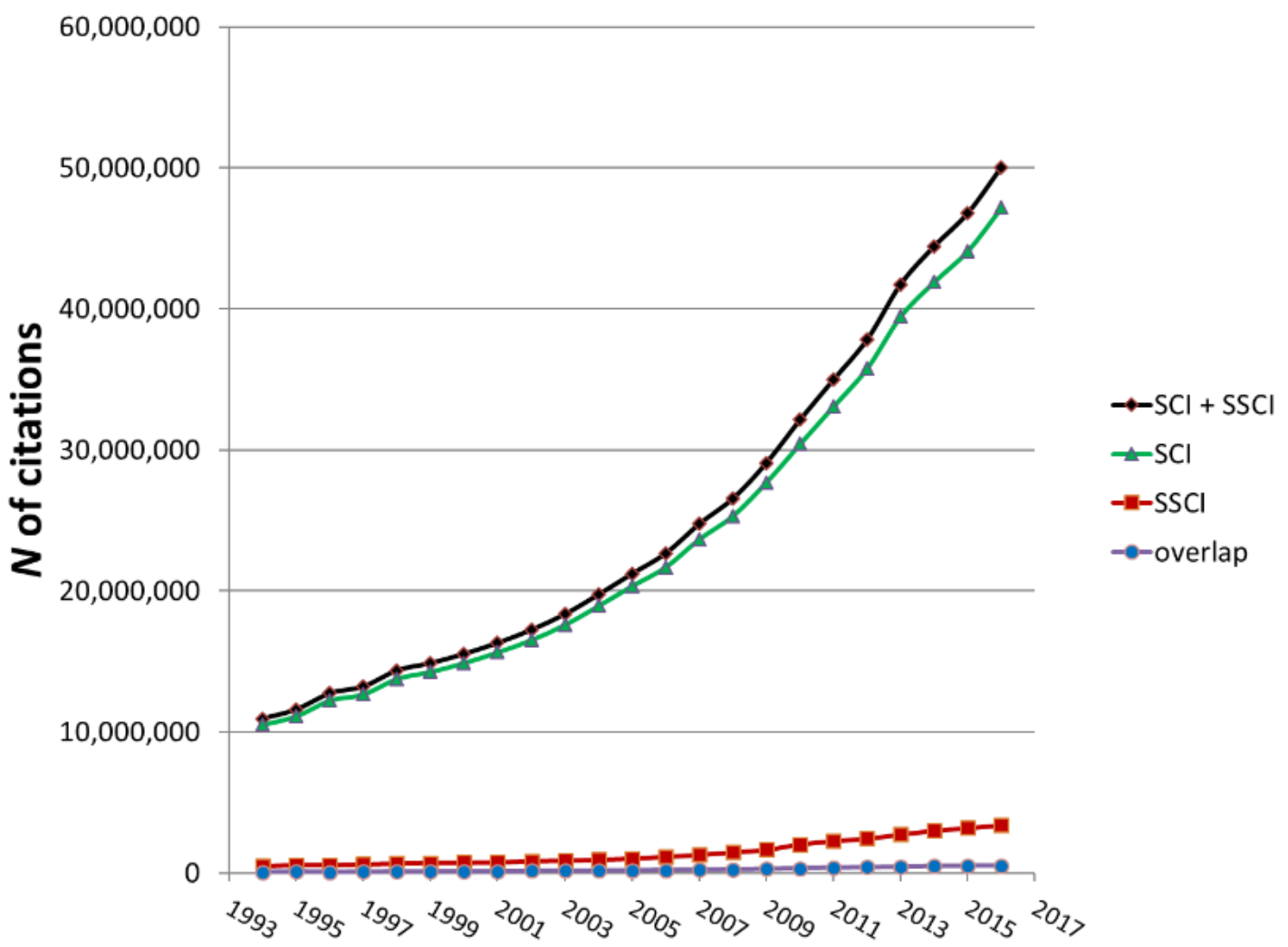

**Figure 2:** Numbers of aggregated citations among journals in WoS for the *Science Citation Index* and the *Social Sciences Citation Index*, both separate and combined.

Figure 1 shows the growth of the SCI and SSCI in terms of the numbers of journals covered during the period under study; Figure 2 adds the numbers of citations stored in the two databases. The difference between SCI and SSCI in the volume of citation is striking. Whereas the contribution of SSCI to the combined set grows from 24.3% of the journals in 1994 to 28.2% in 2016, the citations in the SSCI contribute only 4.6% to the combined set in 1994 and 6.8% in 2016. The overlap in terms of journals grows from 150 (2.6%) in 1994 to 633 (5.5%) in 2016. However, we count these journals only once. Note that the regional expansion of the journal set in 2009 (in Figure 1) did not significantly affect the time series of the cumulative citations in Figure 2.

*2.2. Changes in journal names*

Each *JCR* (in both databases) contains a file with name changes, mergers, and splittings of journals. We use the latest name abbreviation and backtrack from the most recent year of change adding this name to all previous years. For example, the name of the *Journal of Zhejiang University Science C: Computers & Electronics* was changed to *Frontiers of Information Technology & Electronic Engineering* in 2015. We use the abbreviated journal name (FRONT INFORM TECH EL) to follow this journal from its first inclusion into the database in 2011. These uniform labels are needed for comparisons over time.

However, we were not able to backtrack journal names in the case of journal splittings. In such cases, it is not possible to tell which name expresses the line of inheritance in terms of content. Furthermore, journals can be split into more than two new journals. For example, the journal *Biochimica Biophysica Acta* (*BBA*; established in 1947) was split into nine titles as sections of *BBA. BBA – Molecular and Cell Biology of Lipids* commenced in 1998 and is itself a continuation of *BBA – Lipids and Lipid Metabolism* (1965–1998) and *BBA – Specialized Section on Lipids and Related Subjects* (1963–1964). In other words, *BBA* can nowadays be considered as a family of (Elsevier) journals.

In summary, we organize the data of the *JCR*s (*SCI* and *SSCI* combined) into a matrix for each year during the period 1994-2016 with unique identifiers for each cell; that is, the citing and cited journal name abbreviations. The matrix is asymmetrical ("citing" as column variables and "cited" as row variables), but equally sized in both directions (that is, 1-mode). The 23 matrices were thereafter transformed into 21 matrices (1996-2016) with three-year moving averages. Since each evaluation requires three years (*t* as the *a posteriori* year, *t*-1 for the revision of the prediction, and at *t*-2 as the *a priori* distribution), we have 19 observations for each cell (1998-2006). In the final step, we incorporate only values larger than ten (as the aggregate for three years) in order to suppress the possible noise effects of small values.

## 3. Methods

### *3.1. From comparative statics to dynamic analysis*

There are both theoretical and methodological reasons for not using the differences between maps of consecutive years as indicators of change. Cluster analysis and similar techniques (e.g., community finding) are static analyses. Furthermore, community-finding algorithms often begin with a random seed, so that it may even be difficult to reproduce the cluster structure in the same year. When one compares the results of static analyses year-on-year by subtraction, one loses control of whether one is measuring substantive change or the choice of another sub-optimum by the clustering algorithm. One risks confounding the dynamic analysis with the development of error in the measurement as a consequence of the model.

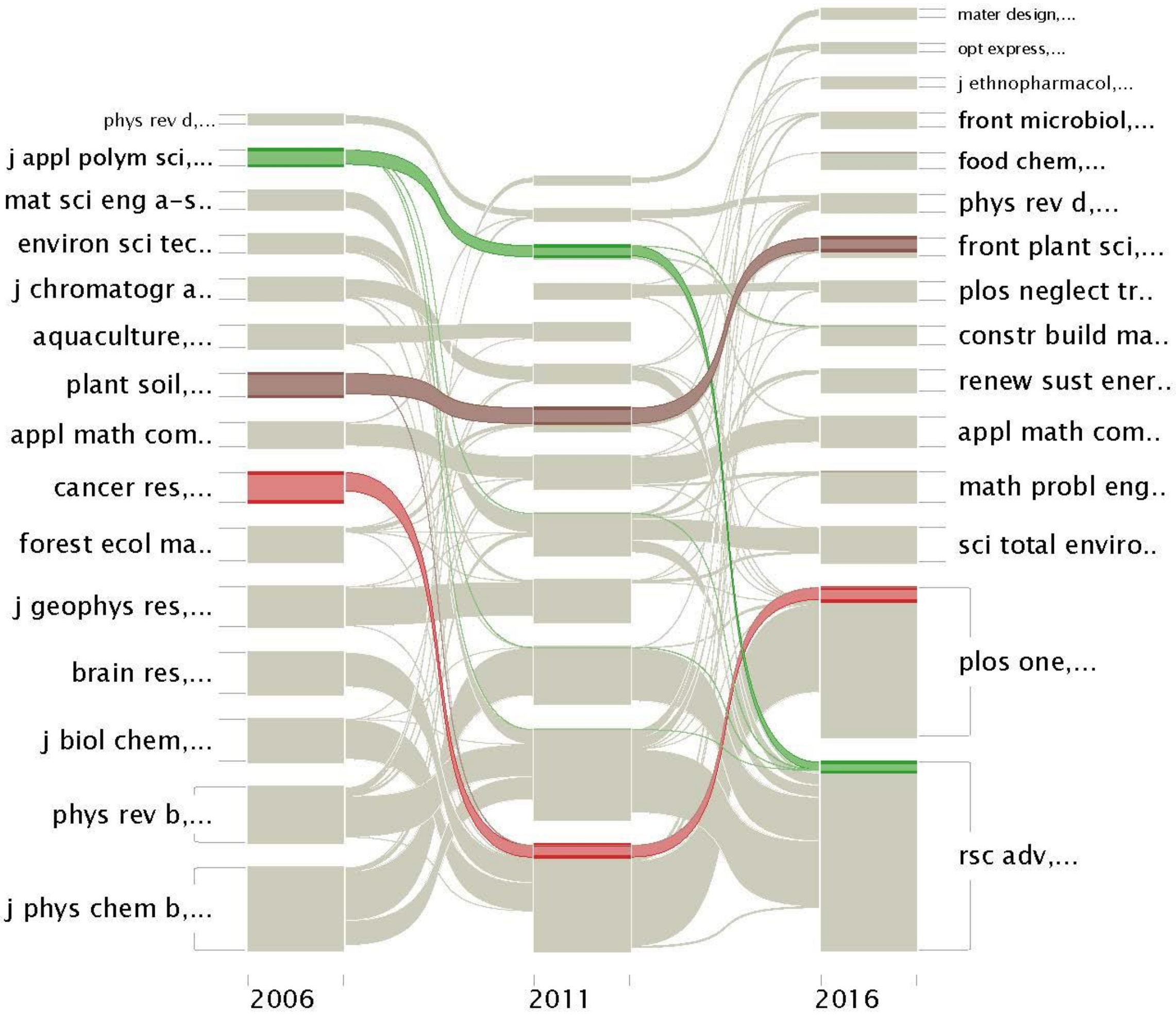

**Figure 1**: Alluvial diagram of three-year moving averages of journal-journal citation relations in 2005, 2011, 2016, using (Rosvall & Bergstrom, 2010)'s *Mapequation* at http://www.mapequation.org/ .

For example, Figure 1 shows an alluvial diagram based on the three-year moving average matrices of journal-journal relations in 2006, 2011, and 2016. It illustrates the problems: (1) journals and journal groups may be positioned differently from year to year because of different (sub)optimalizations in the delineation of clusters, and (2) the emergence of *PLOS ONE* in 2006, and even more importantly the clustering by *RSC Advances* (established by the Royal Society of Chemistry in 2011). Unlike multi-disciplinary journals such as *Science* and *Nature*, these new journals are oriented towards publishing a large volume of papers.

The emergence of these "mega-journals" has made the clustering of journals an unreliable basis for assessing *longitudinal* developments using comparative statics. The ratios of contributions by different specialties or disciplines may change from year to year for reasons very different from substantive ones. This relatively new development adds to the methodological point that comparative static clustering cannot provide a basis for longitudinal inferences in the multi-variate case. Thus, one needs dynamic measures for the longitudinal evaluation.

### *3.2. Change and discontinuity*

Based on Baur & Schank's (2008) dynamic extension of multi-dimensional scaling (MDS), Leydesdorff & Schank (2008) elaborated a dynamic version of *visone* for bibliometric network analysis. However, the size of the networks here under study is computationally beyond the capacity of these routines.

Information theory enables us to study longitudinal developments first at the level of cells and then to aggregate, since the Shannon-formulae are based on using Σs. There are virtually no size limitations for the decomposition of large networks since the evaluation is on a cell-by-cell basis. Note that Rosvall & Bergstrom's (2010) algorithm used information theory for the decomposition and is therefore equally fit for handling (virtually unlimited) large sets. The limits are set only by computer hardware limitations such as available memory and disk size.

The dynamic extension of Shannon's (1948) definition of the information content of a distribution [$H = -\sum_i p_i \log_2 p_i$] is provided by Kullback & Leibler's (1951) divergence measure $I = \Sigma_i\, q_i \log_2 (q_i / p_i)$. where $I$ measures the expected information of the message that the *prior* distribution ($\Sigma_i\, p_i$) has turned into the *posterior* distribution ($\Sigma_i\, q_i$). (When the two-base of the logarithm is used, $I$ is expressed in bits of information.) Theil (1972, at pp. 59 f.) has proven that $I$ is necessarily equal or larger than zero. However, the non-negative *aggregated* value for $I$ allows for local entropy-changes as contributions which are negative. Note furthermore that $I$ is *asymmetrical* in $p$ and $q$: the information content of a change along the arrow of time is different from one in the reverse (backward) direction.

The *prior* distribution can also be considered as a prediction of the *posterior* one. In the case of a perfect prediction, $I = 0$ and the two distributions are similar. If the prediction is imperfect, it can be improved by a distribution at an in-between moment of time (Figure 4). This improvement of the prediction of the *a posteriori* probability distribution ($\Sigma_i\, q_i$) on the basis of an in-between probability distribution ($\Sigma_i\, p_i'$) compared with the original prediction ($\Sigma_i\, p_i$) can be formulated as follows (Theil, 1972, at p. 77):

$$I(q:p) - I(q:p') = \sum_i q_i \log_2 (q_i / p_i) - \sum_i q_i \log_2 (q_i / p_i')$$

$$= \sum_i q_i \log_2 (p_i' / p_i) \qquad (1)$$

If $I(q{:}p) > I(q{:}p') + I(p'{:}p)$, the path via the revision (in Figure 4) is a more efficient channel for the communication between sender and receiver in terms of bits of information than their direct link.

Contrary to the geometry of Figure 4, the sum of the information distances via the intermediate station is then shorter than the direct information path between the sender and the receiver.

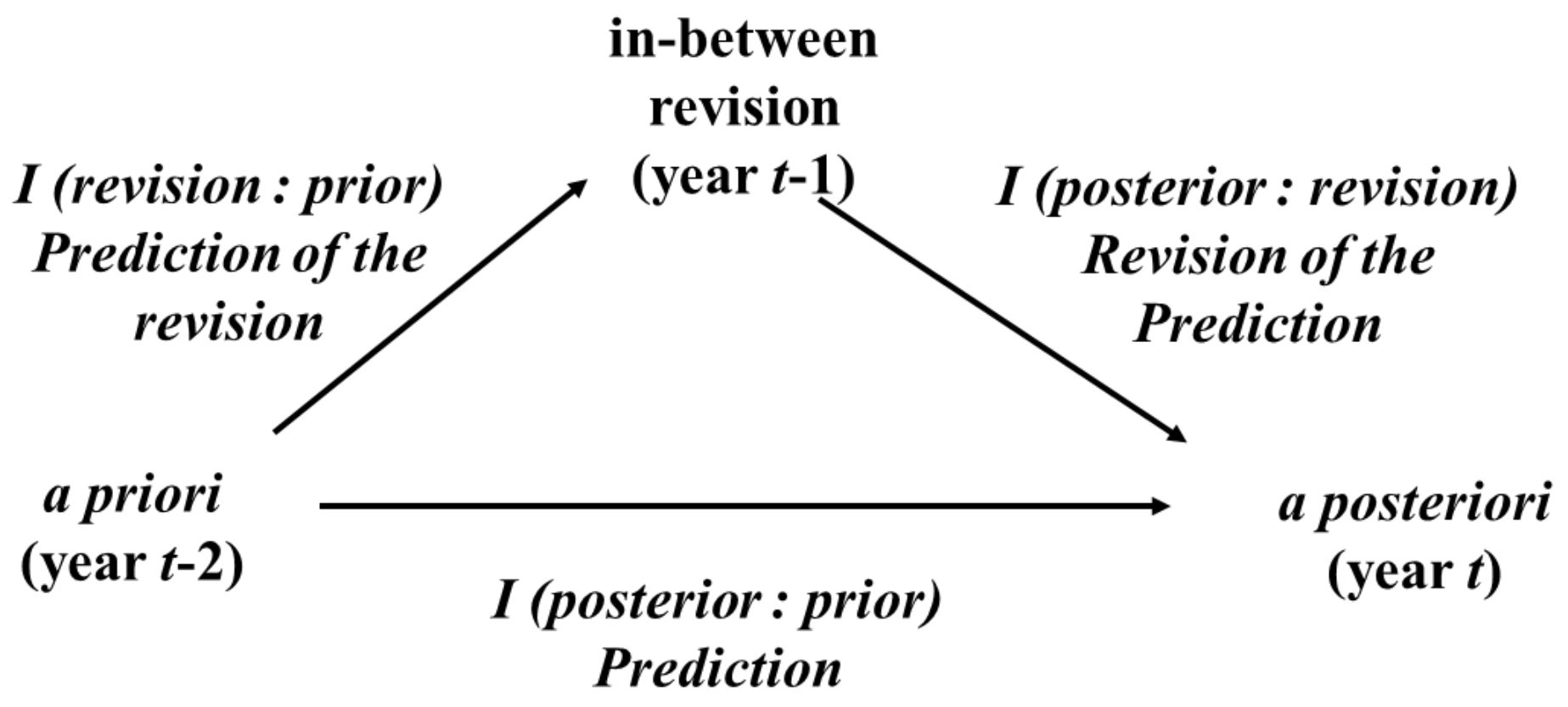

**Figure 4**. Prediction and possible revision of the prediction among three distributions.

The in-between year can also be considered as an auxiliary station in the signal transmission from the sender to the receiver. If the auxiliary station boosts the signal from the sender to the receiver, the system loses its history because what happened before the rewrite no longer matters. In other words, the generation of a negative entropy indicates a discontinuity. The Kullback-Leibler divergences can thus be used to analyze critical or path-dependent transitions in a set of sequential events (Frenken & Leydesdorff, 2000; Leydesdorff, 1991; 1995, at p. 341).

Whereas a path dependency is generated historically if $I(q{:}p') + I(p'{:}p) < I(q{:}p)$ in the forward direction, Lucio-Arias & Leydesdorff (2009) noted that the forward arrow of time models diffusion from a sender to a receiver, whereas the backward arrow assumes the perspective of a receiver looking backward (Rousseau, Zhang, & Hu, in preparation). The backward perspective of the citing receiver provides meaning to the events from the perspective of hindsight. Using the same notation, the inequality can analogously be formulated as: $I(p{:}p') + I(p'{:}q) < I(p{:}q)$.

In summary, let us define an indicator

$$U = I(q{:}p') + I(p'{:}p) - I(q{:}p) \qquad (2)$$

in which $q$ indicates posterior, $p$ prior, and $p'$ revision of the prediction. If $U < 0$, the transition is critical in the forward direction. In other words, negative entropy is generated along the arrow of time. Analogously:

$$V = I(p{:}p') + I(p'{:}q) - I(p{:}q) \qquad (3)$$

indicates a critical transition against the arrow of time (that is, from the perspective of hindsight). The signs in the equations are chosen so that the reasoning remains consistent with (Shannon-type) information theory: a negative entropy indicates the irregularity which needs to be explained. A mathematical elaboration of these equations is provided below in Appendices I and II.

Since $I(p{:}q)$ is unequal to $I(q{:}p)$, one can expect the path dependencies with the arrow of time (representing diffusion) to be different from the ones against the arrow of time (representing codification). While we are interested in the significance of discontinuities for codification more than diffusion, we focus the presentation on the results with the reverse arrows. However, the differences are often small.

## 4. Results

We first explore the data for 2016 (as the most recent year) and then extend the analysis to the full set 1994-2016. In 2016, the domain is based on the ($11,487^2$ =) 131,951,169 possible relations among 11,487 journals. Of these cells, 3,020,242 (2.3%) have a value larger than zero in the *JCR*, containing in total 50,030,365 citation relations. On average, this is 16.6 per cell. However, citation distributions are very skewed: 97.7% of the cells are empty.

In accordance with the findings of de Nooy & Leydesdorff (2015), we found so much volatility when comparing the data for 2016 with the preceding years 2014 and 2015 that we decided to use three-year moving aggregates. A three-year moving aggregate or average, however, is only reliable for relations present during each of three consecutive years (*t*, *t*-1, and *t*-2). In 2016, this is the case for 1,387,423 relations (45.9%) containing 122,778,368 citation relations aggregated over the three years 2014-2016.[8] The average cell value is now (122,778,368 / 3) / 1,387,432 = 29.5. We further reduce noise by using a threshold of above ten in each of the three consecutive years for the assessment of critical transitions and improvements of the prediction. This leads to a file of 844,476 unique journal-journal citation relations[9] that will be used in the analysis of 2016 data, and analogously in the other years.

[8] This is derived as follows: (122,778,368 /3 =) 40,926,123 relations on average, or 81.8% of the original set.
[9] This is 28.0% of the original 3,020,242 non-zero relations and 60.9% of the 1,387,423 cells with three-year moving averages.

*4.1. Discontinuities*

In addition to the *N* of links for each year, Figure 5 shows the longitudinal development of the numbers of observations for which the predictions are improved and path-dependencies are generated during the period 1998-2016. Some of the lines are virtually coinciding: the improvement of the prediction in the forward direction almost coincides with the *absence* of a path-dependency in both the forward and backward directions ($U > 0$ or $V > 0$). This is the case in on average 43.0% of the observations (st. dev. = 1.8%).[10] Secondly, path-dependency in the forward direction coincides with path-dependency in the backward direction and improvement in the backward prediction (57.0% of the observations). We had not expected this coincidence; but we derive in the Appendices why this is the case for analytical reasons.

[10] The improvements have a larger standard deviation (3.0%) than the path-dependencies (1.8%).

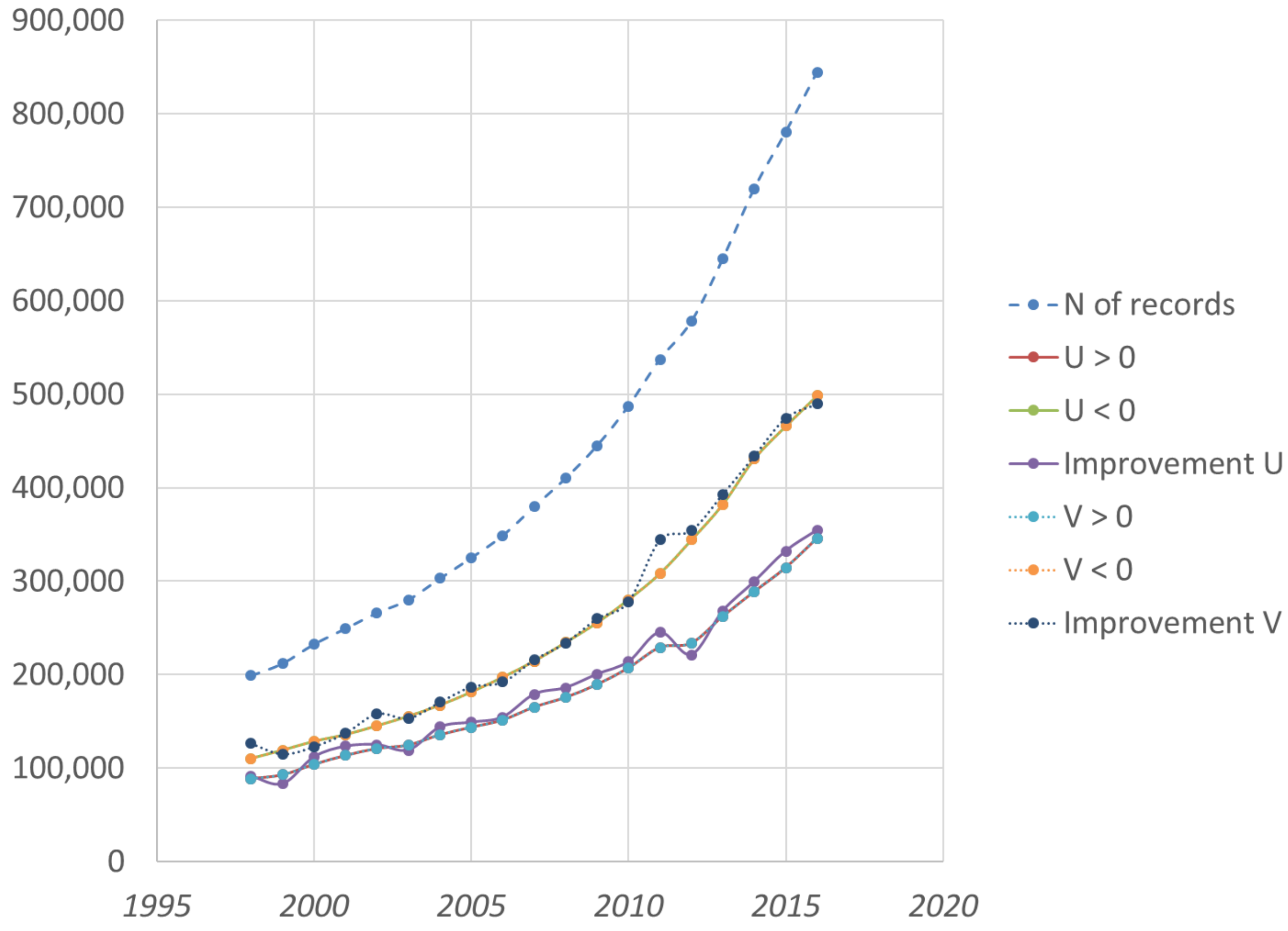

**Figure 5**: Numbers of observations, 1998-2016

The patterns are repeated from year to year. Note that critical transitions are the rule (57%) more than the exception. The values of the critical transitions (in bits), however, vary widely (Figure 6).

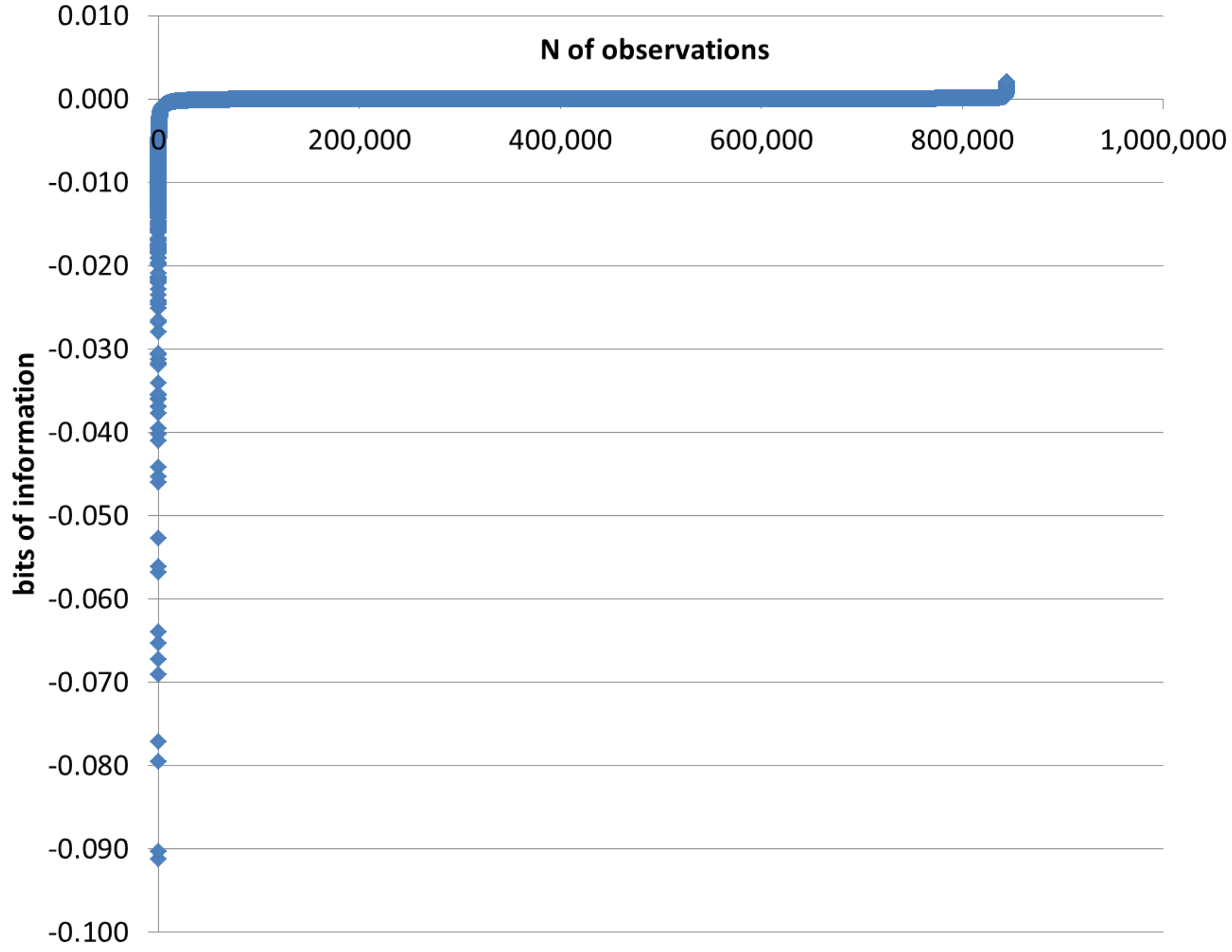

**Figure 6**: Distribution of values of $U = [I(p:p') + I(p':q) - I(p:q)]$ in 2016.

The number of observations in 2016 is 844,476, of which 770,170 (91.2%) have values between -0.1 and +0.1 millibits. This large segment is represented in Figure 6 as a flat line along the *x*-coordinate. At both ends, however, the critical transitions can have much larger absolute values. Plotting these values log-log for the top 10,000 on either end provides two power-law-type distributions (Figure 7) with an excellent fit ($r^2 > .99$).

We added the equations to the figures to show the exponents, which are on the order of 0.7. For a scale-free network, this exponent has to be larger than one (e.g., Broido & Clauset, 2018). The distribution also fits more than .99 to a non-scale-free distribution such as the log-normal or Weibull distributions. The interpretation of this finding is therefore not trivial.

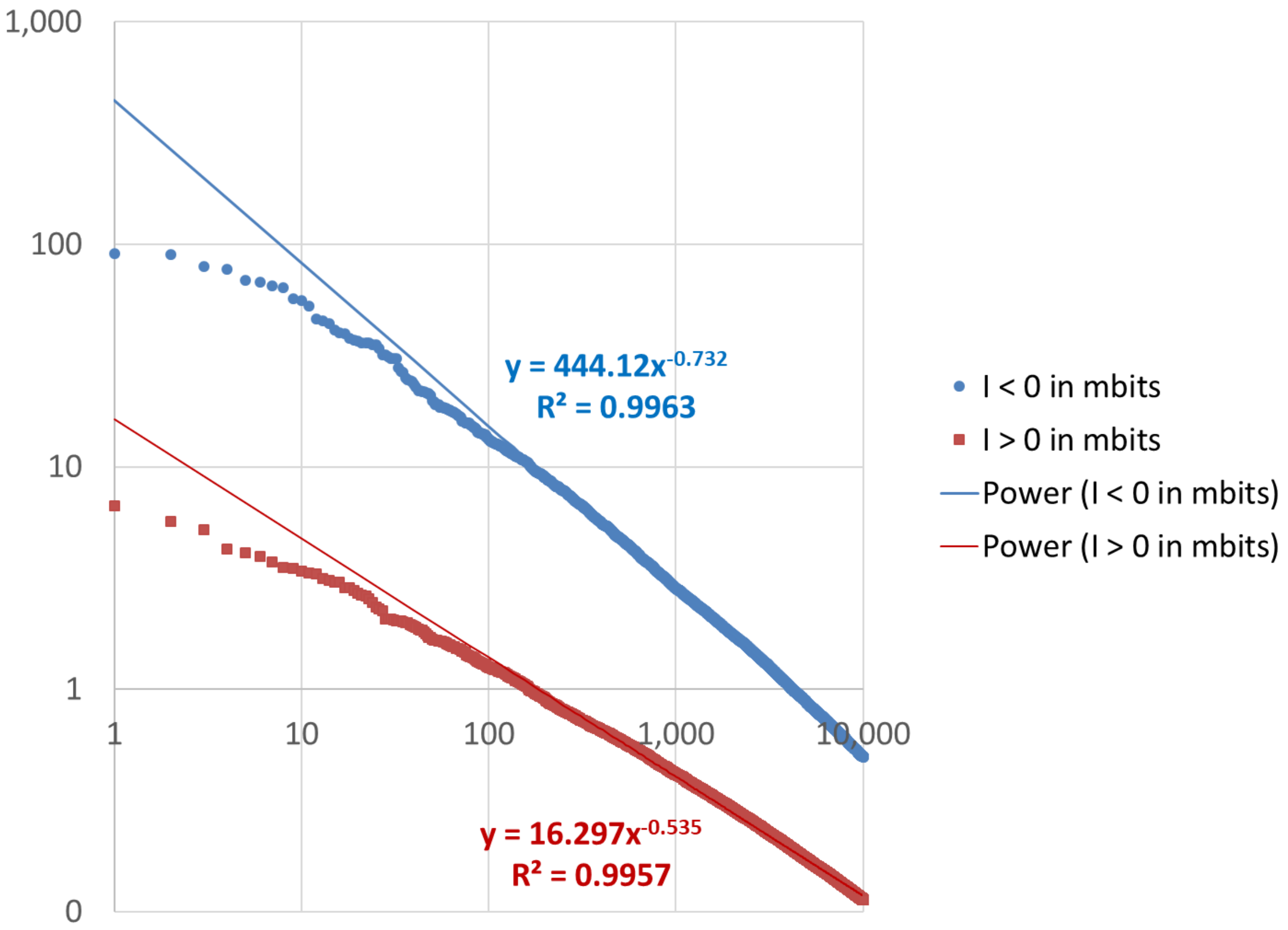

**Figure 7**: The 10,000 most important path-dependent transitions (in blue) versus the 10,000 highest positive values (in brown)

One possible interpretation of the curves might be that the fit marks the signature of self-organized criticality or 1/*f*-noise. For explaining self-organized criticality, Bak & Chen (1991, at pp. 26f.) used the example of a pile of sand on which one grain of sand is dropped regularly: "Now and then, when the slope becomes too steep somewhere on the pile, the grains slide down, causing a small avalanche. […] When a grain of sand is added to a pile in the critical state, it can start an avalanche of any size, including a 'catastrophic' event. But most of the time, the grain will fall so that no avalanche occurs." Even the largest avalanches involve only a small proportion of the grains in the pile, and therefore even catastrophic avalanches cannot cause the slope of the pile to deviate significantly from the critical slope.

In other words, the effects of an avalanche are local and do not affect the overall structure of the pile. The system remains in a critical state so that one can expect avalanches to remain equally possible. "Even though sand is added to the pile at a uniform rate, the amount of sand flowing off the pile varies greatly over time." In contrast to white noise, $1/f$ noise suggests that the dynamics of the system are strongly influenced by past events. The pile has a history of construction and reorganizations over time. Self-organized criticality can also be studied by using a cellular automaton for the simulation.

### *4.2. Self-organized criticality as a model of the development of journal literature*

The fit with a power-law in Figure 7 suggests self-organized criticality (SoC) in the system of journal-journal citation relations. New knowledge claims in manuscripts continuously generate journal-journal citation relations potentially leading to an equivalent of "avalanches" of reconstructed inter-journal relations. These avalanches can occur anywhere; their effects may be very different; but the consequences are local, that is, within the discipline or specialty. The self-organized criticality remains globally available for new critical transitions in the full range from minor, but more frequently occurring changes, to rare but disruptive ones. The comparison with earthquakes provides another metaphor.

This model of self-organized criticality differs from the Kuhnian model of normal science versus revolutionary science as phases in paradigm transitions (Kuhn, 1962; Marx & Bornmann, 2013; van den Daele & Weingart, 1975). The flux of manuscripts with knowledge claims contain references to other journals which can be compared with the grains of sand that hit the sand pile or, in this case, the knowledge base as a construct. The effect can be an avalanche of any size depending on the state of the system at that specific place and time. The selection environments determine the size of the avalanches more than the intrinsic qualities of the knowledge claims providing the variation.

Unlike grains of sand, however, one expects knowledge claims to be related. Bak and his colleagues (Bak & Chen, 1991; Bak, Tang, & Wiesenfeld, 1987) worked in their physical experiments with actual sand grains of uniform granularity. Our "grains" are dropped on a sand pile, but they are of different granularity in that they may be impure, containing, for example, lumps of clay. Golyk (*s. d.*) compared Bak's model with Zipf's Law, which states that in literary texts, the frequency of a word is inversely proportional to its rank in the frequency table, given a large sample of words used. As against sand grains, such texts have complicated non-local correlations such as syntax and cognitive structures (e.g. references), yet the accumulation leads similarly to log-log lines (Price, 1976).

For self-organized criticality to occur, a large number of unrelated meta-stable configurations is needed. Golyk concluded that both "models with local interactions (such as BTW) as well as models with non-local (literary texts) correlations may lead to power-law distributions" (at p. 4).

However, the theory of self-organized criticality has hitherto been rather phenomenological. There is no strict criterion for the value of the exponent, such as one finds in the case of preferential attachment leading to power-law distributions where one uses $2 < \alpha < 3$ as a criterion for scale-freeness. Bak *et al*. (1987, at p. 383) report values of the exponent as low as .42 in studies of SOC. As noted, SOC can be simulated using a cellular automaton as a grid. The exponent is also determined by the dimensionality of the model, and perhaps by the different objects of study such as earthquakes, water droplets on surfaces, human brains, etc. (Jensen, 1998). The original claim that SOC would be scale-free (Bak *et al*., 1987, p. 381) is not needed for SOC as a phenomenon. Scale-free networks are rare (Broido & Clauset, 2018), whereas SOC is abundantly the case in very different systems.

*4.3. Aggregation of the critical links into networks*

Our units of analysis are links, and thus the avalanches also occur in terms of links. Building links upon links, one can expect the link structures to become meta-stable at numerous places. A region may be "poised" for change based upon tension (Foster *et al*., 2015) where tradition and innovation coexist, but "tradition-shattering complements to the tradition-bound activity of normal science" (Kuhn, 1969, p. 6; 1977) may radically reconstruct the organization. Thus, an event introducing new links or abandoning old ones may lead to large, but local shifts at specific places and moments. The non-local correlations hold the structures otherwise together. Sets of journals may be reoriented without losing their cohesive structure as a group.

Moreover, one would not expect multiple disruptions of the system to co-occur. The system operates at the edge of chaos, and thus will respond to perturbations, but an overall system collapse and revolution would not be expected. This suggests that other parts of the network will be robust and resistant to change, even with considerable tensions within them. However, the links that are involved in a reconstruction can again be considered as parts of a network.

Using the value of $U < -1$ mbit as a threshold, for example, 4,207 links are carried by 1,633 journals. Of these journals 1,532 form a single (large) component (Figure 8). Among these journals are not surprisingly multi-disciplinary journals such as *PLOS ONE, PNAS, Science,* and *Nature*. These journals are also part of the citation structures in the contributing disciplines. However, the content of these journals is not subject-bound and varies in its disciplinary composition from year to year. Therefore, they seem to change radically from year to year in terms of the cited knowledge base, but this is an effect of their versatility in terms of disciplinary orientation. However, further away from the center, one finds disciplinarily specific groups of journals and titles indicating specialisms. For example, a cluster focusing on renewable energy sources is positioned on the right side of Figure 8. We return to this cluster below.

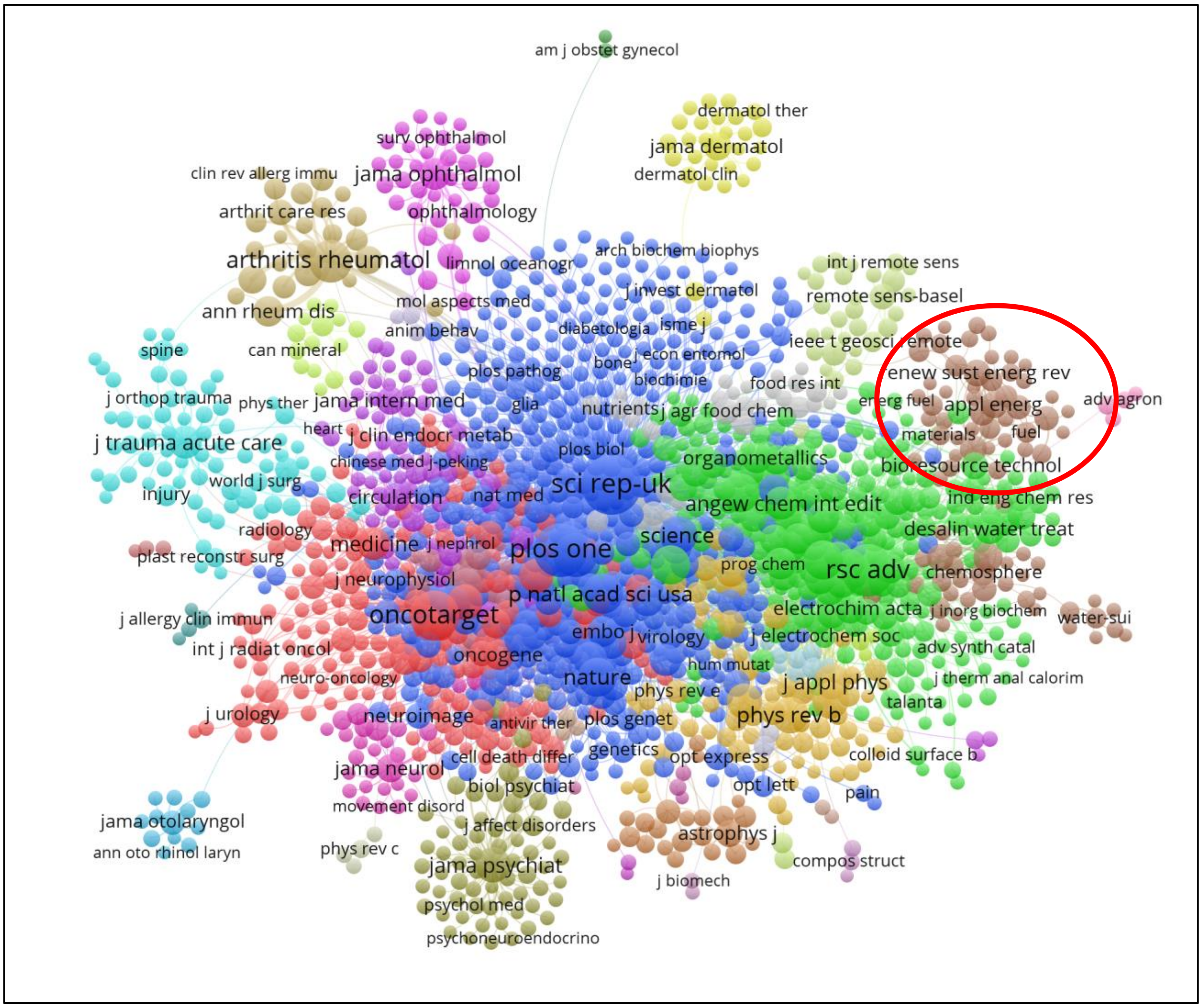

**Figure 8**: Largest component of network among 1,532 journals with a path-dependency of less than -1 mbit in 2016; *N* of Clusters = 37; Modularity *Q* = 0.536.

First, Figure 9 shows forty other components comprising 101 journals and related with values of $U < -1$ mbit, but not related to the main component. However, the threshold of –1 mbit is arbitrary (and set for reasons of presentation); with a lower threshold, more journals and clusters are involved, and with a higher threshold the main component is further decomposed. In other words, one can consider the groups as islands where "avalanches" of various sizes are indicated.

Our parameter choices (e.g., thresholds) structure what we can observe of these avalanches, which can be assumed to occur at all scales and continuously.

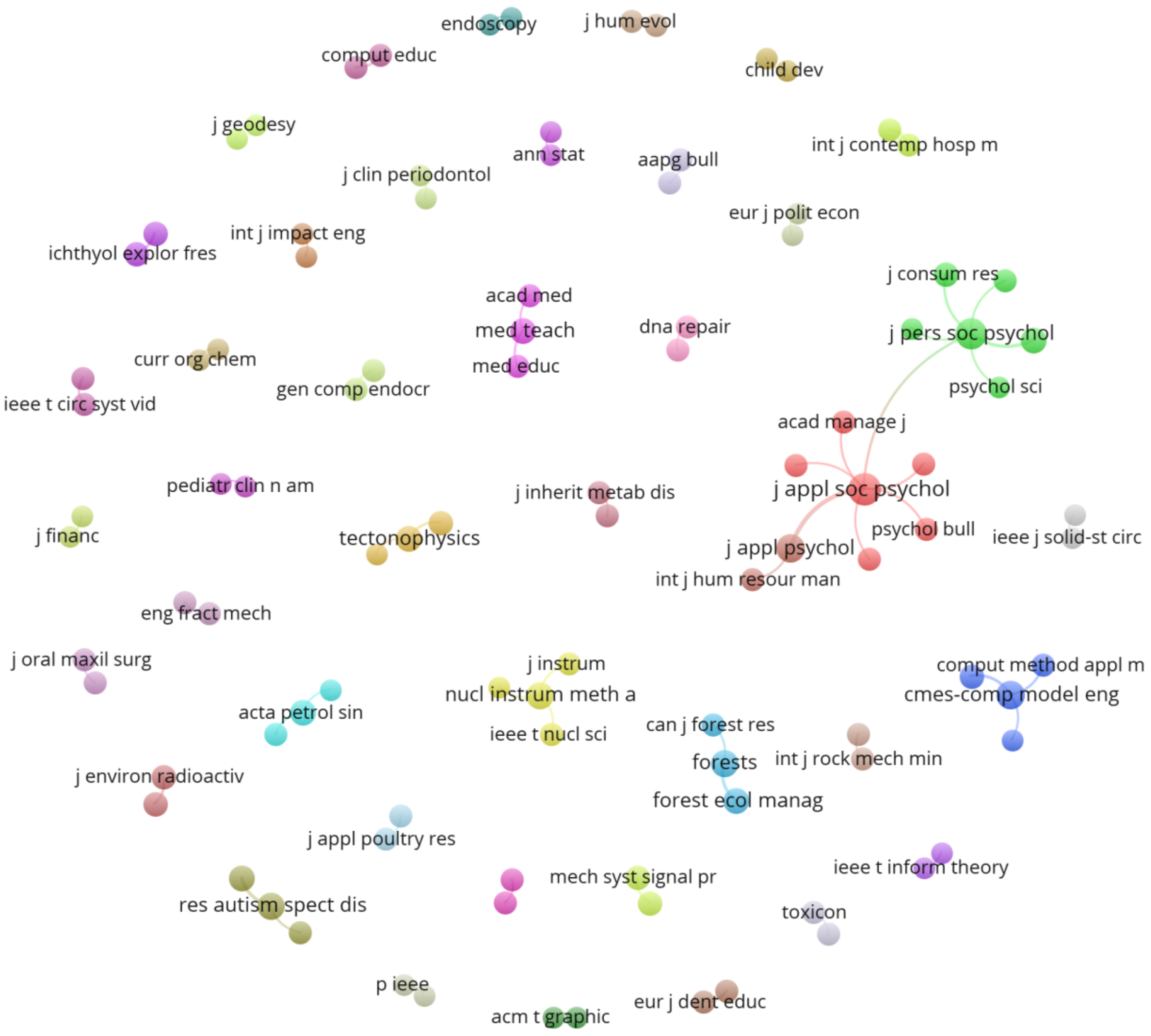

**Figure 9**: 40 components; 101 journals; related at the level of less than -1 mbit.

### *4.3. Changes between years*

In Figure 10, we compare the components which contain the journals *Renewable Energy* and *Renewable and Sustainable Energy Reviews*. In 2010 and 2016, both journals were part of the main component. In 2004—another six years earlier—neither journal was involved in any critical transitions of more than one bit. Using Blondel *et al*.'s (2008) algorithm for community finding, we can extract a subcluster of 72 journals in 2010 and 85 in 2016. The overlap is only 23 journals. How are the non-overlapping journals providing different contexts?

In 2010 (Figure 10A), the group of journals representing this specialty is positioned at the top of the map, but only connected via *Renewable and Sustainable Energy Reviews* to journals focusing on "biomass" as the renewable energy resource at that time. In 2016 (Figure 10B), the group (light blue) is much more central. The focus on biomass has been replaced with one on construction, environmental pollution, and electrical power systems. In Figure 11, the two maps are compared as overlays using dynamic MDS (Leydesdorff & Schank, 2008). The journals with red labels in the upper part show the situation in 2010, and the blue labels at the bottom half the situation in 2016; in the middle is an overlap. Over a period of six years, the orientation of this field has changed.

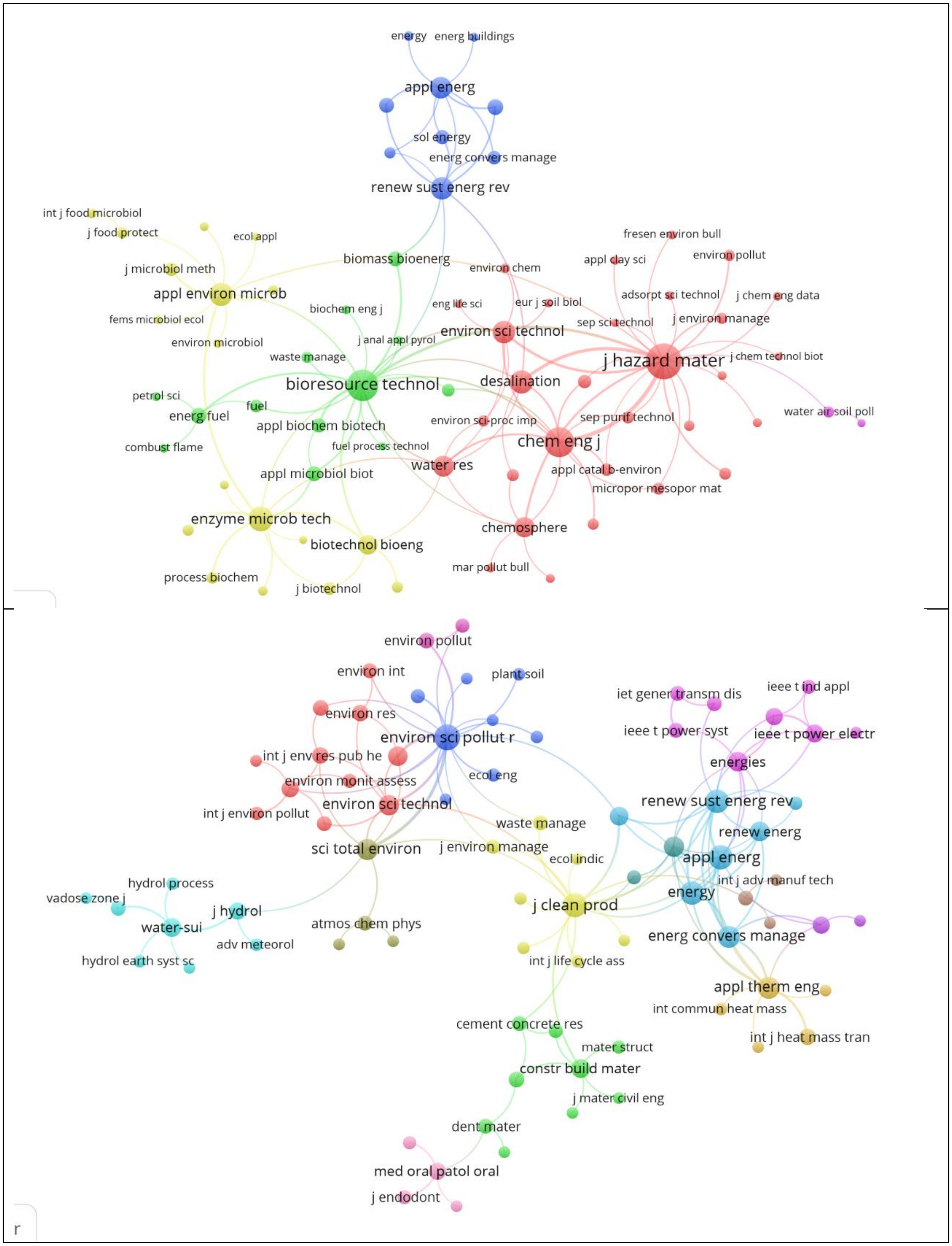

**Figures 10a and 10b**: Cluster of journals about renewable energy in relevant environments in 2010 (Figure 10a at the top) and 2016 (Figure 10b at the bottom), respectively.

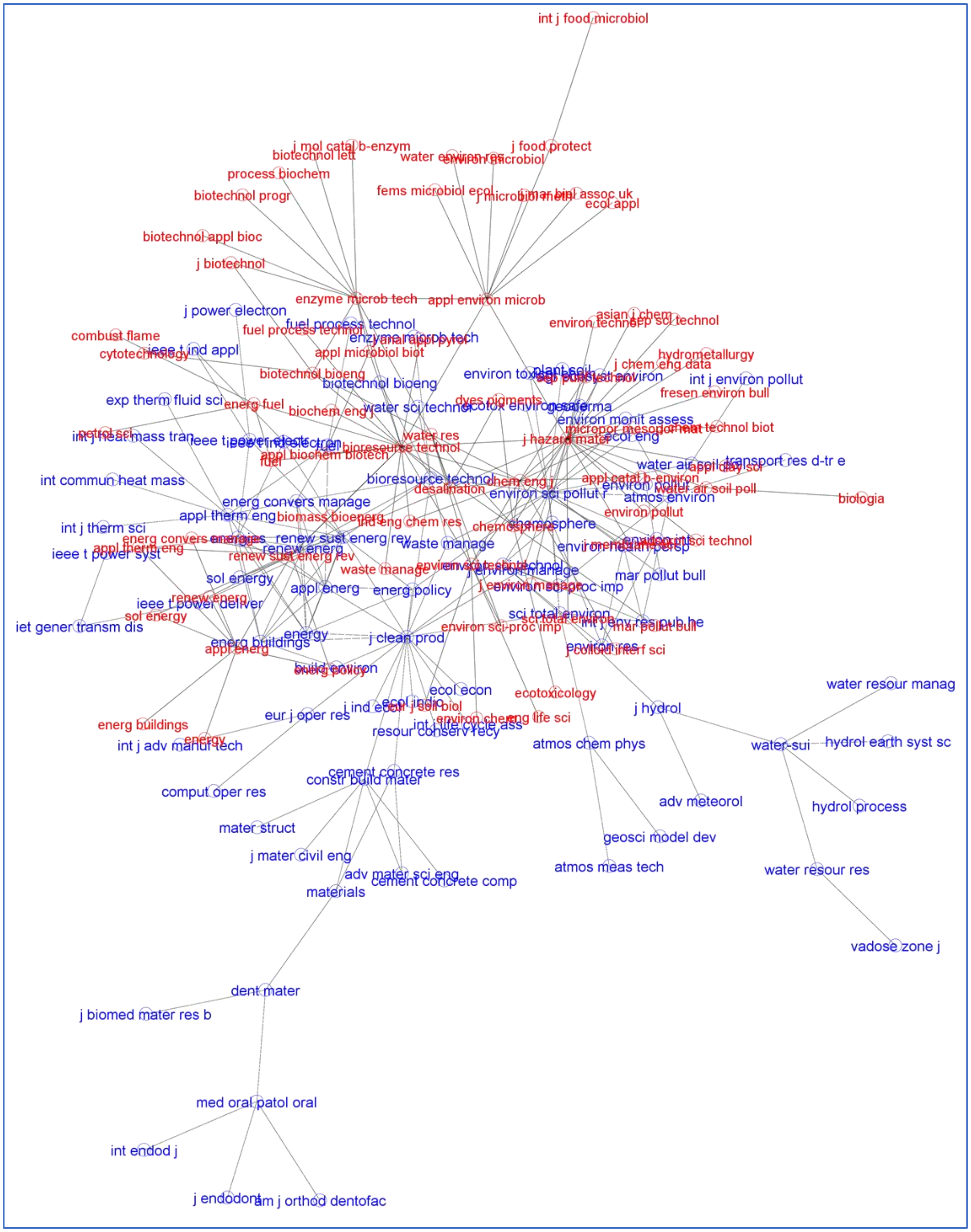

**Figure 11**: Shift of the reseach orientation (citing) between 2010 (red) and 2016 (blue); 72 journals in 2010, 85 in 2016, and 23 journals in the overlap.

*4.4. Testing self-organized criticality*

Since we have 19 observations, we can test whether the power-law in Figure 5 is reproducible in other years. Indeed, we found this fit (*r* > .99) in all years. Figure 12 shows this for the years 1998, 2004, 2010, and 2016. In sum, the phenomenon cam be reproduced.

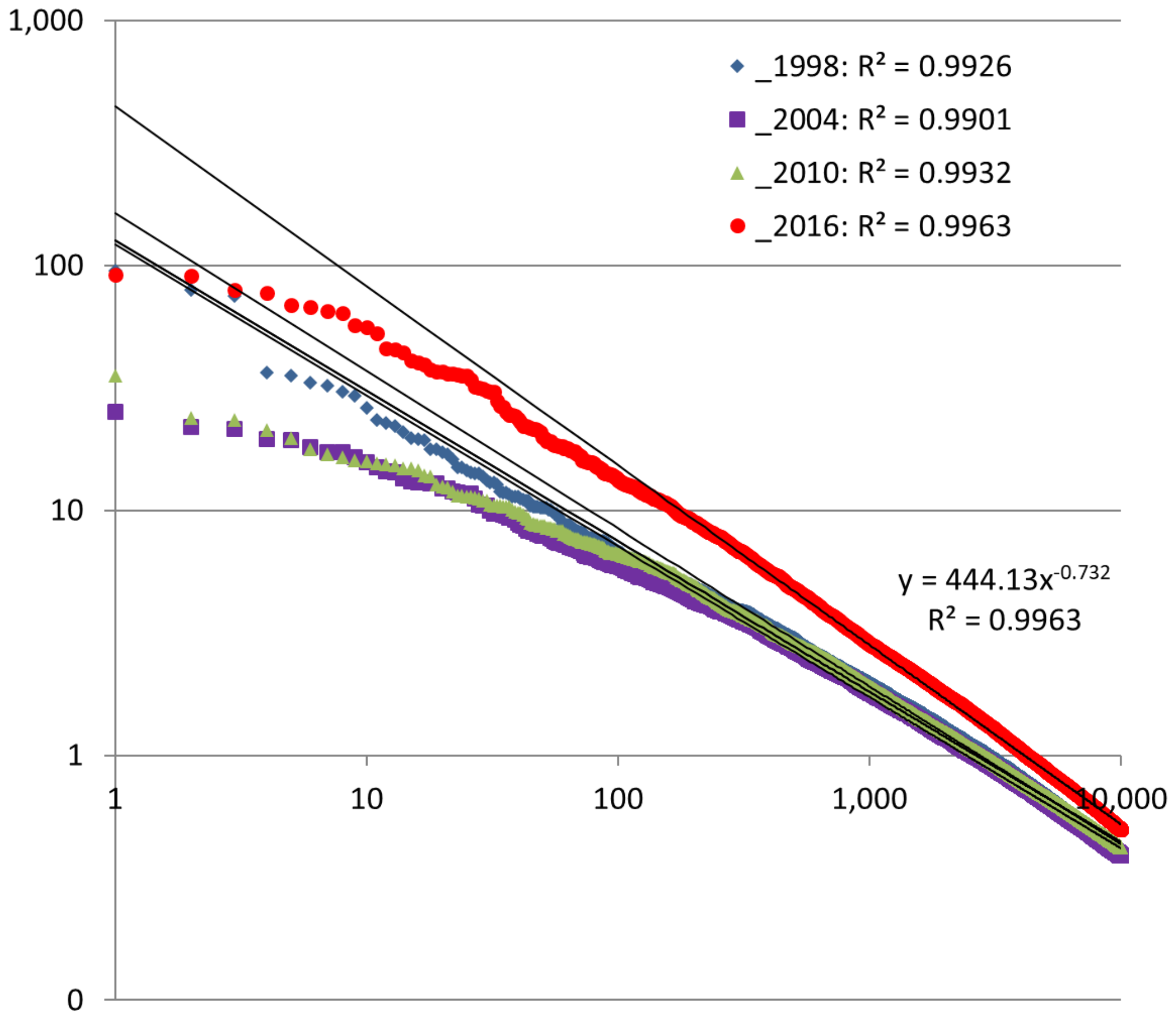

**Figure 12:** Powerlaw distributions of the first 10,000 critical transitions in 1998, 2004, 2010, and 2016.

## 5. Summary and conclusions

Using dynamic entropy measures such as Kullback-Leibler's (1951) divergence, Theil's (1972) improvement of the prediction, and Leydesdorff's (1991) test for critical transitions as indicators of change over time, we unexpectedly found "self-organized criticality" in the 10,000+ most pronounced cases of evolutionary change. It seems to us that the model of self-organized criticality makes it possible to show regularities that help us to understand the evolutionary development of the sciences where disciplines are both enabled and constrained by what is past—or conventional—and what is possible or can be sustained by the system. While it is impossible to say where change will occur, the fact that it will occur is expected. Smaller, local events may be fractals of what occurs elsewhere on a larger scale. However, we do not expect the system to be scale-free, since the scaling coefficients may vary among disciplines.

The sciences can be understood as developing in terms of interdependent continuities and discontinuities. The continuities are needed for the accumulation of the "sand pile" into a knowledge base, with the inertia of institutionalized relations maintaining the relevant structures. The discontinuities provide options for a rewrite (Fujigaki, 1998) or a reorganization. These may happen as lightning ("grains of sand") striking the ground—in an unpredictable way—but the effect can be a local "avalanche" of any size and accordingly a reorganization.

Unlike a sand pile, the knowledge base is a construct comprising specific structures. Whereas some multidisciplinary journals participate in this reorganization of the knowledge base every year—because of their multidisciplinary character—large numbers of journal links are unresponsive to changes in the environment, so that the transitions remain close to zero in terms of bits of information. The system is both dynamic by chance and structurally "frozen," but in different parts of the structure. Kauffman & Johnson (1991) described this as a co-evolution to the edge of chaos; the constructed knowledge base is at many places meta-stable, while at other places, the systems may be temporarily locked and therefore unable to change.

In addition to this unexpected result, the use of critical transitions enables us to show how a research front can shift rapidly. In the case of research about renewable energy, for example, the research focus on "biomass" in 2010 shifted to an orientation towards themes like electrical power, construction, and clean production in 2016. The journals were present in 2004, but this research area was not yet involved in the dynamics of discontinuity.

The data allow for raising many more questions, for example, about the dynamics of disciplinary groups of journals (e.g., Leydesdorff & Schank, 2008; Rosvall & Bergstrom, 2010). It might be interesting to follow up with a study of the social sciences separately. Another extension would be a further study about codification on the citing side versus the diffusion of a signal from a source to a set of receivers—as is a common assumption in citation analysis (cf. Zitt & Small, 2008). As noted, what events mean is decided from the perspective of hindsight using deterministic (!) selection mechanisms that operate on the forward generation of variation. One obvious limitation of the present study is the assumption that citations are simple counts, independent of the questions of who is citing, what is cited, at which place, in which context, etc.

**Acknowledgement**

The authors wish to thank Clarivate for making the *JCR* data available.

**Appendix I**: Elaboration of the inequalities in the case of critical transitions.

A transition is critical in the forward direction if:

$$q * \log_2 {}^{q}/_{p} > q * \log_2 {}^{q}/_{p'} + p' * \log_2 {}^{p'}/_{p}$$

Or after elaboration of the logarithms as exponents:

$$\frac{q^q}{p^q} > \frac{q^q}{(p')^q} * \frac{p'^{p'}}{p^{p'}}$$

$$p^{-q} > (p')^{-q} * (p')^{p'} * (p)^{-p'}$$

$$(p)^{p'-q} > (p')^{p'-q}$$

if ($p' > q$ then $p > p' > q$) OR (if $p' < q$ then $p < p' < q$)

Analogously in the backward direction:

$$p * \log_2 {}^{p}/_{q} > p * \log_2 {}^{p}/_{p'} + p' * \log_2 {}^{p'}/_{q}$$

$$\frac{p^p}{q^p} > \frac{p^p}{(p')^p} * \frac{p'^{p'}}{q^{p'}}$$

$$q^{-p} > (p')^{-p} * (p')^{p'} * (q)^{-p'}$$

$$(q)^{p'-p} > (p')^{p'-p}$$

(if $p' > p$ then $q > p' > p$) OR ( if $p' < p$ then $q < p' < p$)

These conditions are the same in both directions: when the sequence between the probabilities increases or decreases monotonically (with or against the arrow of time), the transition is path-dependent.

**Appendix II**: Elaboration of the inequalities in the case of improvements of the prediction.

The condition for an improvement of the prediction in the forward direction is:

$$q \log_2 {p'}/{p} > 0$$

$$\frac{(p\prime)^q}{(p)^q} > 0$$

This is true if $p' > p$. In the other case ($p' < p$), the perdiction is worsened.

*Mutatis mutandis* in the backward direction:

$$\frac{(p\prime)^p}{(q)^p} > 0$$

This is true if $p' > q$. In the other case ($p' < q$), the prediction is worsened.

As against the symmetrical conditions for critical revisions, improvement is asymmetrical in both the backward and the forward direction.